\documentclass[aip,jcp,reprint,groupedaddress,floatfix]{revtex4-1}

\usepackage{amsmath}
\usepackage{amssymb}
\usepackage{dsfont}
\usepackage{graphicx}
\usepackage{mathtools}
\usepackage[version=4]{mhchem}
\usepackage{microtype}
\usepackage{physics}
\usepackage{siunitx}

\usepackage[hidelinks]{hyperref}

\renewcommand{\vec}[1]{\boldsymbol{\mathbf{#1}}}

\renewcommand{\epsilon}{\varepsilon}
\renewcommand{\phi}{\varphi}
\renewcommand{\rho}{\varrho}

\newcommand{\inv}{^{-1}}
\newcommand{\st}{^\star}
\newcommand{\cl}{_\mathrm{cl}}
\newcommand{\ddf}[1]{\delta\qty(#1)}
\newcommand{\dlogJ}[1]{\pdv{\xi} \log{\abs*{J(#1)}}}

\newcommand{\estim}{\mathcal{E}}

\begin{document}

\title{
	On the quantum mechanical potential of mean force. I. A path integral perspective
}

\author{Dmitri Iouchtchenko}
\author{Kevin P. Bishop}
\author{Pierre-Nicholas Roy}
\email{pnroy@uwaterloo.ca}
\affiliation{Department of Chemistry, University of Waterloo, Waterloo, Ontario, N2L 3G1, Canada}

\begin{abstract}
We derive two path integral estimators for the derivative of the quantum mechanical potential of mean force (PMF), which may be numerically integrated to yield the PMF.
For the first estimator, we perform the differentiation on the exact path integral, and for the second, we perform the differentiation on the path integral after discretization.
These estimators are successfully validated against reference results for the harmonic oscillator and Lennard-Jones dimer systems using constrained path integral Monte Carlo (PIMC) simulations.
Specifically, the estimators reproduce both the derivative of the PMF, as well as the PMF itself, for the model systems at multiple temperatures.
In Paper II, these estimators are implemented alongside path integral molecular dynamics (PIMD) with a constrained path integral Langevin equation thermostat for use with more general systems and potentials.
\end{abstract}

\maketitle

\section{Introduction}
\label{sec:introduction}

Free energy calculations provide vital theoretical insights into the behaviour of chemical systems (such as equilibrium constants and stability of molecular conformations) and are commonly used to make comparisons with experimental results.
The classical potential of mean force (PMF), also referred to as a free energy profile, may be obtained from classical molecular dynamics simulations in a number of ways: umbrella sampling with the weighted histogram analysis method (WHAM),\cite{torrie1977nonphysical,kumar1992weighted,roux1995calculation} blue moon sampling,\cite{carter1989constrained,sprik1998free,ciccotti2005blue} metadynamics,\cite{laio2002escaping,bussi2006equilibrium} and potential of mean constraint force,\cite{mulders1996free,den2000free} among others.

The quantum mechanical analogue of the PMF has not been as thoroughly studied, but there are many interesting systems where nuclear quantum effects play an integral role.
These effects are essential within simulations performed at low temperature or containing light atoms.
Crucially, the inclusion of nuclear quantum effects has been demonstrated to be indispensable in the accurate determination of properties of the water dimer as well as other small water clusters due to the presence of the light hydrogen atoms.\cite{ceriotti2016nuclear,markland2018nuclear,vaillant2018path}
Recent work has combined the existing umbrella sampling method with path integral molecular dynamics simulations to study the free energy profile of water--water\cite{bishop2018quantum,mendez2020nuclear} and water--methanol dimers\cite{mendez2020nuclear} while accounting for such nuclear quantum effects.

Many existing efforts to compute the quantum PMF\cite{hinsen1997potential,major2006path,major2007integrated,walker2010direct,vardi2012hybrid} have used the path integral centroid coordinate.
The so-called centroid PMF can be defined both in the path integral representation\cite{cao1994formulation1,cao1994formulation4,voth96,nb2001I} and in the operator\cite{jang99I,jang99II,roy99I,roy99II,reichman2000,nb2001II,nb2002I,orr2017formulation} formulations of centroid statistical mechanics.
The centroid is a path property and does not correspond to a physical observable.
This fact is known to produce marked deviations from the exact quantum mechanical result.\cite{blinov2004connection,bishop2018quantum}
For instance, the use of the centroid radial distribution function and its associated structure factor has led to the wrong interpretation of scattering experiments on liquid para-hydrogen.\cite{bermejo2000quantum,blinov2004connection}
Our current focus is therefore to obtain the PMF as a function of a physical observable, the true reaction coordinate, rather than the centroid, a path property.

In the present work, we formally derive two path integral Monte Carlo (PIMC) estimators for the derivative of the PMF, which can be integrated to determine the PMF.
The first estimator is obtained by performing an analytical differentiation of the exact path integral, followed by its discretization over the path.
Conversely, the second estimator is derived by initially discretizing the path integral before performing the analytical differentiation.
Theoretically, these estimators should provide the same numerical results, and to verify this, they are benchmarked against known model systems.

Both estimators may be used in conjunction with the path integral Langevin equation (PILE),\cite{ceriotti2010efficient} as we show in Paper II of this series, titled ``Constrained path integral molecular dynamics integrators''.
Together with the constrained PILE formulation, we aim to use these estimators for systems that are not as easily studied with PIMC, such as low-temperature molecular clusters.

The remainder of this article is organized as follows: in Sec.~\ref{sec:background}, we describe our notation; in Sec.~\ref{sec:estimators}, we develop two estimators for the derivative of the PMF; in Sec.~\ref{sec:results}, we apply the estimators to model systems; and in Sec.~\ref{sec:conclusions}, we summarize our findings.

\section{Background}
\label{sec:background}

We consider systems with $f$ Cartesian degrees of freedom, that we label $q_1$, $q_2$, \dots, $q_f$; commonly, there are $N$ particles in three spatial dimensions, in which case $f = 3 N$.
For convenience, we group them into a single vector $\vec{q}$.

We restrict the Hamiltonian to have the form
\begin{align}
	\label{eq:hamiltonian}
	\hat{H}
	&= \hat{K} + \hat{V}
	= \sum_{i=1}^f \frac{\hat{p}_i^2}{2 m_i} + V(\hat{\vec{q}})
	= \frac{1}{2} \hat{\vec{p}} \cdot \vec{M}\inv \cdot \hat{\vec{p}} + V(\hat{\vec{q}}),
\end{align}
where $\hat{p}_i$ is the momentum operator conjugate to the position operator $\hat{q}_i$, $m_i$ is the mass corresponding to $q_i$, and $\vec{M}$ is the diagonal mass matrix whose elements are $m_i$.
The restriction on the kinetic energy allows us to write the exact free particle propagator
\begin{align}
	\mel{\vec{q}'}{e^{-\tau \hat{K}}}{\vec{q}}
	&= \sqrt{\frac{\abs{\vec{M}}}{(2 \pi \hbar^2 \tau)^f}} \, e^{-\frac{1}{2 \hbar^2 \tau} (\vec{q}' - \vec{q}) \cdot \vec{M} \cdot (\vec{q}' - \vec{q})}
\end{align}
for an imaginary time duration $\tau$.
We require that the potential energy be diagonal in the position representation so that
\begin{align}
	\mel{\vec{q}'}{e^{-\tau \hat{V}}}{\vec{q}}
	&= \ddf{\vec{q}' - \vec{q}} e^{-\tau V(\vec{q})}.
\end{align}
Despite these limitations, such Hamiltonians are general enough to describe many diverse systems of itinerant particles.

The partition function of a system with Hamiltonian $\hat{H}$ at reciprocal temperature $\beta = 1 / k_\mathrm{B} T$ is
\begin{align}
	\label{eq:Z}
	Z
	&= \Tr e^{-\beta \hat{H}}
	= \int\! \dd{\vec{q}} \mel{\vec{q}}{e^{-\beta \hat{H}}}{\vec{q}},
\end{align}
and the thermal expectation value of an operator $\hat{O}$ is
\begin{align}
	\label{eq:expval-quantum}
	\ev*{\hat{O}}_{\beta \hat{H}}
	&= \frac{1}{Z} \Tr e^{-\beta \hat{H}} \hat{O}
	= \frac{
			\int\! \dd{\vec{q}} \mel{\vec{q}}{e^{-\beta \hat{H}} \hat{O}}{\vec{q}}
		}{
			\int\! \dd{\vec{q}} \mel{\vec{q}}{e^{-\beta \hat{H}}}{\vec{q}}
		}.
\end{align}
As a means of evaluating $\ev*{\hat{O}}_{\beta \hat{H}}$, we may construct a discretized imaginary time path integral for the partition function.
To that end, we first rename $\vec{q}$ to $\vec{Q}^{(1)}$ and then insert $P-1$ resolutions of the identity
\begin{align}
	\hat{\mathds{1}}
	&= \int\! \dd{\vec{Q}^{(j)}} \dyad*{\vec{Q}^{(j)}},
\end{align}
which introduce the additional Cartesian coordinates $\vec{Q}^{(2)}$, \dots, $\vec{Q}^{(P)}$ along the imaginary time path; we combine them all into the vector $\vec{Q}$ and refer to them as ``beads'', picturing the path as a necklace.\cite{feynmanhibbs,feynmanstat,Chandler:1981p6533}
This results in
\begin{align}
	Z
	&= \int\! \dd{\vec{Q}} \, \prod_{j=1}^P \mel*{\vec{Q}^{(j)}}{e^{-\frac{\beta}{P} \hat{H}}}{\vec{Q}^{(j+1)}},
\end{align}
where it should be understood that the path is cyclic in imaginary time (that is, $\vec{Q}^{(P+1)}$ is an alias for $\vec{Q}^{(1)}$).

To evaluate each high-temperature propagator, since $\comm*{\hat{K}}{\hat{V}} \ne 0$, we rely on the Trotter factorization
\begin{align}
	e^{-\beta \hat{H}}
	&= \lim_{P \to \infty} \left( e^{-\frac{\beta}{P} \hat{K}} e^{-\frac{\beta}{P} \hat{V}} \right)^P,
\end{align}
which allows us to start with the approximation
\begin{align}
	\mel{\vec{q}'}{e^{-\frac{\beta}{P} \hat{H}}}{\vec{q}}
	&\approx\! \sqrt{\frac{\abs{\vec{M}} P^f}{(2 \pi \hbar^2 \beta)^f}} \,
			e^{
				-\frac{P}{2 \hbar^2 \beta} (\vec{q}' - \vec{q}) \cdot \vec{M} \cdot (\vec{q}' - \vec{q})
				- \frac{\beta}{P} V(\vec{q})
			}
\end{align}
and systematically improve the error in the product of these approximate factors by increasing $P$.
For any finite $P$, we may construct the approximate path density
\begin{align}
	\pi(\vec{Q})
	&= \left( \frac{\abs{\vec{M}} P^f}{(2 \pi \hbar^2 \beta)^f} \right)^{\frac{P}{2}}
			e^{-\beta V\cl(\vec{Q})}
\end{align}
with the classical potential
\begin{align}
	V\cl(\vec{Q})
	&= \sum_{i=1}^f \frac{m_i P}{2 \hbar^2 \beta^2} \sum_{j=1}^P \left( Q^{(j)}_i - Q^{(j+1)}_i \right)^2
	\notag \\ &\qquad
		+ \frac{1}{P} \sum_{j=1}^P V(\vec{Q}^{(j)}),
\end{align}
so that
\begin{align}
	Z
	&= \lim_{P \to \infty} \int\! \dd{\vec{Q}} \, \pi(\vec{Q}).
\end{align}
For the remainder of this article, we drop the $P \to \infty$ limit for the sake of brevity.

In order to use PIMC sampling to calculate $\ev*{\hat{O}}_{\beta \hat{H}}$, it is necessary to procure an estimator function $\mathcal{E}_{\hat{O}}(\vec{Q})$, the details of which depend on the nature of the operator.
The operator expression in Eq.~\eqref{eq:expval-quantum} is then replaced by a ratio of integrals containing only regular functions:
\begin{align}
	\ev*{\hat{O}}_{\beta \hat{H}}
	&= \ev*{\mathcal{E}_{\hat{O}}}_{\pi}
	= \frac{
			\int\! \dd{\vec{Q}} \, \pi(\vec{Q}) \mathcal{E}_{\hat{O}}(\vec{Q})
		}{
			\int\! \dd{\vec{Q}} \, \pi(\vec{Q})
		}.
\end{align}
This ratio is commonly evaluated as
\begin{align}
	\ev*{\mathcal{E}_{\hat{O}}}_{\pi}
	&\approx \frac{1}{N_\mathrm{MC}} \sum_{i=1}^{N_\mathrm{MC}} \mathcal{E}_{\hat{O}}(\vec{Q}_{[i]})
\end{align}
by drawing the samples $\{ \vec{Q}_{[i]} \}_{i=1}^{N_\mathrm{MC}}$ from $\pi(\vec{Q})$ using Markov chain Monte Carlo.

\section{Estimators}
\label{sec:estimators}

It is generally more convenient to work with path integrals in Cartesian coordinates $\vec{q}$, but the PMF $A(\xi\st)$ is expressed in terms of an arbitrary curvilinear coordinate $\xi$ at some value $\xi\st$.
To connect the two, we introduce an invertible coordinate transformation to the generalized coordinates $X_1$, $X_2$, \dots, $X_{f-1}$, $\xi$, where the first $f - 1$ of these are grouped into the vector $\vec{X}$.
This transformation has non-zero Jacobian determinant $J(\vec{q}) = J(\vec{X}, \xi)$.
The special coordinate $\xi$ is referred to as the \emph{reaction coordinate}; for example, it may be the distance between two specific centers of mass in a cluster.

Using the diagonal reduced density
\begin{subequations}
\label{eq:density-generalized}
\begin{align}
	\rho(\xi\st)
	&= \frac{1}{Z} \mel{\xi\st}{\Tr_{\vec{X}} e^{-\beta \hat{H}}}{\xi\st} \\
	&= \frac{1}{Z} \int\! \dd{\vec{X}} \mel{\vec{X} \, \xi\st}{e^{-\beta \hat{H}}}{\vec{X} \, \xi\st}
\end{align}
\end{subequations}
at reciprocal temperature $\beta$, we may construct the overall object of interest: the potential of mean force
\begin{align}
	A(\xi\st)
	&= -\frac{1}{\beta} \log{\frac{\rho(\xi\st)}{\rho_0}},
\end{align}
where $\rho_0$ is an arbitrary constant with the same physical dimension as $\xi^{-1}$.
Choosing a value for $\rho_0$ sets the zero of energy for the PMF.
Note that our definitions imply that $\rho_0^{-1} = \int\! \dd{\xi\st} e^{-\beta A(\xi\st)}$, which does not contain any explicit volume element factors; instead, we encounter a geometric term in the estimators.
Even though the momentum operator $\hat{p}_\xi$ conjugate to the reaction coordinate operator $\hat{\xi}$ satisfies\cite{leaf1980curvilinear}
\begin{align}
	\bra{\xi} \hat{p}_\xi
	&= -i \hbar \pdv{\xi} \bra{\xi},
\end{align}
in general we find that
\begin{align}
	\bra{\vec{q}} \hat{p}_\xi
	&\ne -i \hbar \pdv{\xi} \bra{\vec{q}},
\end{align}
and the missing portion is directly responsible for the geometric term.

We wish to compute $A(\xi\st)$ via its derivative
\begin{align}
	A'(\xi\st)
	&= -\frac{1}{\beta} \dv{\xi\st} \log{\frac{\rho(\xi\st)}{\rho_0}}
	= -\frac{1}{\beta} \frac{\rho'(\xi\st)}{\rho(\xi\st)}.
\end{align}
As shown in Appendix~\ref{sec:curvilinear}, we may write the diagonal reduced density in Cartesian coordinates as
\begin{align}
	\rho(\xi\st)
	&= \frac{1}{Z} \int\! \dd{\vec{q}} \, \ddf{\xi(\vec{q}) - \xi\st} \mel{\vec{q}}{e^{-\beta \hat{H}}}{\vec{q}},
\end{align}
so
\begin{align}
	-\beta A'(\xi\st)
	&= \frac{
			\dv{\xi\st} \int\! \dd{\vec{q}} \, \ddf{\xi(\vec{q}) - \xi\st} \mel{\vec{q}}{e^{-\beta \hat{H}}}{\vec{q}}
		}{
			\int\! \dd{\vec{q}} \, \ddf{\xi(\vec{q}) - \xi\st} \mel{\vec{q}}{e^{-\beta \hat{H}}}{\vec{q}}
		}.
\end{align}
Because the denominator resembles a constrained version of the partition function $Z$ in Eq.~\eqref{eq:Z}, we use this as the starting point to derive two path integral estimators $\mathcal{E}_1(\vec{Q})$ and $\mathcal{E}_2(\vec{Q})$ which satisfy
\begin{align}
	-\beta A'(\xi\st)
	&= \!\!\ev*{\mathcal{E}_i}_{\pi,\xi\st}\!
	= \frac{
			\int\! \dd{\vec{Q}} \, \ddf{\xi(\vec{Q}^{(1)}) - \xi\st} \pi(\vec{Q}) \mathcal{E}_i(\vec{Q})
		}{
			\int\! \dd{\vec{Q}} \, \ddf{\xi(\vec{Q}^{(1)}) - \xi\st} \pi(\vec{Q})
		}.
\end{align}
For $\mathcal{E}_1(\vec{Q})$, we first differentiate and then discretize the path integral in the numerator, and for $\mathcal{E}_2(\vec{Q})$ we do the reverse.

\subsection{Estimator 1: Differentiate then discretize}

The main quantity in question is
\begin{align}
	Z \rho'(\xi\st)
	&= \dv{\xi\st} \int\! \dd{\vec{q}} \, \ddf{\xi(\vec{q}) - \xi\st} \mel{\vec{q}}{e^{-\beta \hat{H}}}{\vec{q}},
\end{align}
which we write using Appendix~\ref{sec:delta-derivative} as
\begin{align}
	\label{eq:density-derivative}
	Z \rho'(\xi\st)
	&= \int\! \dd{\vec{q}} \, \ddf{\xi(\vec{q}) - \xi\st} \left[ J_\xi(\vec{q}) + \pdv{\xi} \right] \mel{\vec{q}}{e^{-\beta \hat{H}}}{\vec{q}}.
\end{align}
The simpler of the two terms is the geometric one, which stems from the coordinate transformation:
\begin{align}
	\label{eq:geometric}
	\int\! \dd{\vec{q}} \, \ddf{\xi(\vec{q}) - \xi\st} \mel{\vec{q}}{e^{-\beta \hat{H}}}{\vec{q}} J_\xi(\vec{q}).
\end{align}
The remaining term
\begin{align}
	\label{eq:non-geometric}
	\int\! \dd{\vec{q}} \, \ddf{\xi(\vec{q}) - \xi\st} \sum_{i=1}^f \pdv{q_i}{\xi} G_i(\vec{q})
\end{align}
is more involved, requiring the derivatives of the imaginary time propagator:
\begin{align}
	G_i(\vec{q})
	&= \pdv{q_i} \mel{\vec{q}}{e^{-\beta \hat{H}}}{\vec{q}}
	= \frac{1}{i \hbar} \mel{\vec{q}}{\comm{e^{-\beta \hat{H}}}{\hat{p}_i}}{\vec{q}},
\end{align}
where the derivative--commutator identity is derived in Appendix~\ref{sec:derivative-commutator}.

Using the Kubo formula for the commutator with the exponential of an operator,\cite{wilcox1967exponential,kubo92} we find that
\begin{align}
	G_i(\vec{q})
	&= -\frac{1}{i \hbar} \int_0^\beta\! \dd{\lambda} \mel{\vec{q}}{e^{-(\beta - \lambda) \hat{H}} \comm*{\hat{H}}{\hat{p}_i} e^{-\lambda \hat{H}}}{\vec{q}}.
\end{align}
Since the kinetic energy operator in Eq.~\eqref{eq:hamiltonian} commutes with $\hat{p}_i$, only the commutator with the potential energy remains: $\comm*{\hat{V}}{\hat{p}_i}$.
It follows from $\comm{\hat{q}_i}{\hat{p}_j} = i \hbar \delta_{i j}$ that
\begin{align}
	\comm{V(\hat{\vec{q}})}{\hat{p}_i}
	&= -i \hbar F_i(\hat{\vec{q}}),
\end{align}
where the component of the force vector $\vec{F}(\vec{q})$ on the coordinate $q_i$ is given by
\begin{align}
	F_i(\vec{q})
	&= -\pdv{q_i} V(\vec{q}).
\end{align}
Hence,
\begin{align}
	G_i(\vec{q})
	&= \int_0^\beta\! \dd{\lambda} \mel{\vec{q}}{e^{-(\beta - \lambda) \hat{H}} F_i(\hat{\vec{q}}) e^{-\lambda \hat{H}}}{\vec{q}}.
\end{align}

Having obtained the necessary expressions, we may discretize the path integral in the usual fashion.
The geometric term in Eq.~\eqref{eq:geometric} poses no difficulty, and we get
\begin{align}
	\label{eq:geometric-discretized}
	\int\! \dd{\vec{Q}} \, \ddf{\xi(\vec{Q}^{(1)}) - \xi\st} \pi(\vec{Q}) J_\xi(\vec{Q}^{(1)}).
\end{align}
The integral from the Kubo formula is discretized into an average over the path, and because $F_i(\hat{\vec{q}})$ is diagonal in the additional path coordinates, we only need to perform the substitution
\begin{align}
	G_i(\vec{q})
	&\to \pi(\vec{Q}) \frac{\beta}{P} \sum_{j=1}^P F_i(\vec{Q}^{(j)}),
\end{align}
turning Eq.~\eqref{eq:non-geometric} into
\begin{align}
	\int\! \dd{\vec{Q}} \, \ddf{\xi(\vec{Q}^{(1)}) - \xi\st} \pi(\vec{Q}) \frac{\beta}{P} \sum_{j=1}^P \vec{F}(\vec{Q}^{(j)}) \cdot \pdv{\vec{Q}^{(1)}}{\xi}.
\end{align}
Thus, the PMF derivative may be written as
\begin{align}
	-\beta A'(\xi\st)
	\! &= \!\!\ev*{\mathcal{E}_1}_{\pi,\xi\st}\!
	= \frac{
			\int\! \dd{\vec{Q}} \, \ddf{\xi(\vec{Q}^{(1)}) - \xi\st} \pi(\vec{Q}) \estim_1(\vec{Q})
		}{
			\int\! \dd{\vec{Q}} \, \ddf{\xi(\vec{Q}^{(1)}) - \xi\st} \pi(\vec{Q})
		},
\end{align}
where
\begin{align}
	\label{eq:estim1}
	\estim_1(\vec{Q})
	&= \dlogJ{\vec{Q}^{(1)}}
		+ \frac{\beta}{P} \sum_{j=1}^P \vec{F}(\vec{Q}^{(j)}) \cdot \pdv{\vec{Q}^{(1)}}{\xi}
\end{align}
is the first estimator.
In the $P = 1$ case, it reduces to a form recognizable from classical mechanics:\cite{den2000free}
\begin{subequations}
\begin{align}
	-\frac{1}{\beta} \estim_1(\vec{q})
	&= -\frac{1}{\beta} \pdv{\xi} \log{\abs*{J(\vec{q})}}
		- \vec{F}(\vec{q}) \cdot \pdv{\vec{q}}{\xi} \\
	&= \pdv{\xi} \left[ V(\vec{q}) - \frac{1}{\beta} \log{\abs*{J(\vec{q})}} \right].
\end{align}
\end{subequations}

\subsection{Estimator 2: Discretize then differentiate}

To derive another estimator, we start from Eq.~\eqref{eq:density-derivative}, but first discretize the path into $P$ imaginary time steps to find
\begin{align}
	Z \rho'(\xi\st)
	&= \int\! \dd{\vec{Q}} \, \ddf{\xi(\vec{Q}^{(1)}) - \xi\st}
	\notag \\ &\qquad\times
			\left[ J_\xi(\vec{Q}^{(1)}) + \pdv{\vec{Q}^{(1)}}{\xi} \cdot \pdv{\vec{Q}^{(1)}} \right] \pi(\vec{Q}).
\end{align}
The geometric term will again be as in Eq.~\eqref{eq:geometric-discretized}, but the other term is now straightforward to compute via ordinary calculus, requiring only
\begin{widetext}
\begin{align}
	\frac{1}{\pi(\vec{Q})} \pdv{\pi(\vec{Q})}{\vec{Q}^{(j)}}
	&= -\beta \pdv{V\cl(\vec{Q})}{\vec{Q}^{(j)}}
	= \beta \vec{F}\cl^{(j)}(\vec{Q})
	= \frac{\beta}{P} \vec{F}(\vec{Q}^{(j)})
		- \frac{P}{\hbar^2 \beta} \vec{M} \cdot \left[ 2 \vec{Q}^{(j)} - \vec{Q}^{(j+1)} - \vec{Q}^{(j-1)} \right],
\end{align}
\end{widetext}
in which the classical force $\vec{F}\cl^{(j)}(\vec{Q})$ on bead $j$ is obtained from the classical potential.
The PMF derivative may therefore also be written as
\begin{align}
	-\beta A'(\xi\st)
	\! &= \!\!\ev*{\mathcal{E}_2}_{\pi,\xi\st}\!
	= \frac{
		\int\! \dd{\vec{Q}} \, \ddf{\xi(\vec{Q}^{(1)}) - \xi\st} \pi(\vec{Q}) \estim_2(\vec{Q})
		}{
			\int\! \dd{\vec{Q}} \, \ddf{\xi(\vec{Q}^{(1)}) - \xi\st} \pi(\vec{Q})
		},
\end{align}
where
\begin{align}
	\label{eq:estim2}
	\estim_2(\vec{Q})
	&= \dlogJ{\vec{Q}^{(1)}}
		+ \beta \vec{F}\cl^{(1)}(\vec{Q}) \cdot \pdv{\vec{Q}^{(1)}}{\xi}
\end{align}
is the second estimator.

It is perhaps a little surprising that the sum $\frac{\beta}{P} \sum_{j=2}^P \vec{F}(\vec{Q}^{(j)})$ from $\mathcal{E}_1$, which involves all coordinates except the constrained one, appears to have been replaced by $-\frac{P}{\hbar^2 \beta} \vec{M} \cdot \left( 2 \vec{Q}^{(1)} - \vec{Q}^{(2)} - \vec{Q}^{(P)} \right)$, which depends on only three coordinates.
This results in the peculiar identity
\begin{align}
	\ev{\vec{F}\cl^{(1)}(\vec{Q}) \cdot \pdv{\vec{Q}^{(1)}}{\xi}}_{\pi,\xi\st}
	\!\!\!\!\!\!\!\!\!&\overset{P \to \infty}{=}\! \ev{\frac{1}{P} \sum_{j=1}^P \vec{F}(\vec{Q}^{(j)}) \cdot \pdv{\vec{Q}^{(1)}}{\xi}}_{\pi,\xi\st},
\end{align}
which relates the classical force on the constrained coordinates to the average force over the path.

\subsection{Removal of geometric term}

It is occasionally more convenient to work with
\begin{align}
	\tilde{\rho}(\xi\st)
	&= \frac{\rho(\xi\st)}{f(\xi\st)},
\end{align}
for some function $f$, than with $\rho(\xi\st)$ directly.
Consider, for example, the spherical coordinates ($\xi$, $\cos{\theta}$, $\phi$), which have Jacobian determinant
\begin{align}
	J(\xi, \cos{\theta}, \phi)
	&= -\xi^2.
\end{align}
The normalization
\begin{align}
	\label{eq:spherical-normalization}
	1
	&= \int\! \dd{\xi\st} (\xi\st)^2 \tilde{\rho}(\xi\st)
\end{align}
is often more natural than
\begin{align}
	1
	&= \int\! \dd{\xi\st} \rho(\xi\st).
\end{align}

Using $\tilde{\rho}$ leads to the modified PMF
\begin{align}
	\tilde{A}(\xi\st)
	&= -\frac{1}{\beta} \log{\frac{\tilde{\rho}(\xi\st)}{\tilde{\rho}_0}}
	= A(\xi\st)
		+ \frac{1}{\beta} \log{\frac{\tilde{\rho}_0 f(\xi\st)}{\rho_0}},
\end{align}
where the arbitrary constant $\tilde{\rho}_0$ has the same physical dimension as $(\xi f(\xi))^{-1}$.
The corresponding derivative is
\begin{align}
	-\beta \tilde{A}'(\xi\st)
	&= -\beta A'(\xi\st)
		- \pdv{\xi} \log{f(\xi\st)},
\end{align}
and it follows immediately that
\begin{subequations}
\begin{align}
	\tilde{\estim}_1(\vec{Q})
	&= \estim_1(\vec{Q}) - \pdv{\xi} \log{f(\xi\st)}
\intertext{and}
	\tilde{\estim_2}(\vec{Q})
	&= \estim_2(\vec{Q}) - \pdv{\xi} \log{f(\xi\st)}
\end{align}
\end{subequations}
may be used to estimate $-\beta \tilde{A}'(\xi\st)$.

Whenever a transformation from $\vec{q}$ to $\vec{X}, \xi$ exists with Jacobian determinant $J(\vec{X}, \xi)$ that is a function of only $\xi$ (as in the above spherical coordinates example) the geometric term may be exactly cancelled from $\estim_1$ and $\estim_2$ by setting $f(\xi) = \abs{J(\xi)}$.
In such a situation,
\begin{align}
	\pdv{\xi} \log{f(\xi\st)}
	&= \dlogJ{\vec{Q}^{(1)}},
\end{align}
so we are left with just
\begin{subequations}
\begin{align}
	\tilde{\estim}_1(\vec{Q})
	&= \frac{\beta}{P} \sum_{j=1}^P \vec{F}(\vec{Q}^{(j)}) \cdot \pdv{\vec{Q}^{(1)}}{\xi}
\intertext{and}
	\tilde{\estim_2}(\vec{Q})
	&= \beta \vec{F}\cl^{(1)}(\vec{Q}) \cdot \pdv{\vec{Q}^{(1)}}{\xi}.
\end{align}
\end{subequations}
This modification provides no substantial computational benefits, as the omitted expression will be a constant with respect to the integration (for example, $2 / \xi\st$ for spherical coordinates).
However, it may make sense to exclude the term from the calculation entirely if it is destined to be excised after the calculation is completed.

\subsection{Kubo formula in generalized coordinates}

Starting from Eq.~\eqref{eq:density-generalized}, which expresses the diagonal reduced density in terms of the generalized coordinates, application of Appendix~\ref{sec:derivative-commutator} immediately yields
\begin{align}
	Z \rho'(\xi\st)
	&= \frac{1}{i \hbar} \int\! \dd{\vec{X}} \mel{\vec{X} \, \xi\st}{\comm{e^{-\beta \hat{H}}}{\hat{p}_\xi}}{\vec{X} \, \xi\st}.
\end{align}
This bypasses many of the convoluted steps found above and leaves us with a succinct expression, which takes on the form
\begin{align}
	\label{eq:generalized-kubo}
	-\frac{1}{i \hbar} \int\! \dd{\vec{X}} \int_0^\beta\! \dd{\lambda} \mel{\vec{X} \, \xi\st}{e^{-(\beta - \lambda) \hat{H}} \comm*{\hat{H}}{\hat{p}_\xi} e^{-\lambda \hat{H}}}{\vec{X} \, \xi\st}
\end{align}
after treatment with the Kubo formula.\cite{wilcox1967exponential}
Since
\begin{align}
	\comm{V(\hat{\vec{q}})}{\hat{p}_\xi}
	&= -i \hbar F_\xi(\hat{\vec{q}})
\end{align}
involves only the force along the reaction coordinate, proceeding in this direction seems like the obvious choice.
However, $\hat{p}_\xi$ is not guaranteed to commute with the Cartesian momenta, and the commutator $\comm*{\hat{K}}{\hat{p}_\xi}$ is not always diagonal in the position representation.
Consequently, Eq.~\eqref{eq:generalized-kubo} does not lend itself well to discretization and we do not pursue this approach to the derivation further.

\section{Results}
\label{sec:results}

As a proof of concept, we use the estimators $\estim_1$ and $\estim_2$ to compute derivatives of the PMF of two small systems, for which reference results (either exact or numerical) may be calculated: a one-dimensional harmonic oscillator and a Lennard-Jones model of the \ce{Ar2} dimer.
To perform the path integral Monte Carlo sampling, we have implemented a basic Markov chain integrator\cite{snaggednecklacejl} using the Metropolis--Hastings acceptance criterion.
The constraint is exactly enforced by sampling in the generalized coordinates for the first bead: updates are proposed for $\vec{X}^{(1)}$, but $\xi^{(1)}$ is held fixed at $\xi\st$.

\subsection{Harmonic oscillator}

The simplest non-trivial problem we can consider is the dependable harmonic oscillator, with the Hamiltonian
\begin{align}
	\hat{H}
	&= \frac{\hat{p}^2}{2 m} + \frac{1}{2} m \omega^2 \hat{q}^2
\end{align}
and the reaction coordinate $\xi = q$.
Since the eigenstates of this Hamiltonian are known analytically, we may write down the normalized diagonal density
\begin{align}
	\rho(\xi\st)
	&= \frac{1}{Z} e^{-\frac{\beta \hbar \omega}{2}} \frac{\alpha}{\sqrt{\pi}} e^{-(\alpha \xi\st)^2}
			\sum_{n=0}^\infty \frac{e^{-\beta \hbar \omega n}}{2^n n!} H_n^2(\alpha \xi\st),
\end{align}
where $H_n(x)$ is the order-$n$ Hermite polynomial at $x$, $\alpha = \sqrt{m \omega / \hbar}$, and the partition function is
\begin{align}
	Z
	&= \frac{1}{2} \csch(\beta \hbar \omega / 2).
\end{align}
Using the identity\cite{hermitesum}
\begin{align}
	\sum_{n=0}^\infty \frac{k^n}{2^n n!} H_n(x) H_n(y)
	&= \frac{e^\frac{k^2 (x^2 + y^2) - 2 k x y}{k^2 - 1}}{\sqrt{1 - k^2}}
\end{align}
for $\abs{k} < 1$, which in our case simplifies to
\begin{align}
	\sum_{n=0}^\infty \frac{k^n}{2^n n!} H_n^2(x)
	&= \frac{e^\frac{2 k x^2}{k + 1}}{\sqrt{1 - k^2}},
\end{align}
we find that
\begin{align}
	\rho(\xi\st)
	&= \sqrt{\frac{\alpha^2 \tanh(\beta \hbar \omega / 2)}{\pi}} e^{-\alpha^2 \tanh(\beta \hbar \omega / 2) (\xi\st)^2}.
\end{align}
Thus,
\begin{align}
	\label{eq:exact-ho1d}
	-\beta A'(\xi\st)
	&= -2 \alpha^2 \tanh(\beta \hbar \omega / 2) \xi\st,
\end{align}
which is proportional to $\xi\st$.

\begin{figure}
	\includegraphics{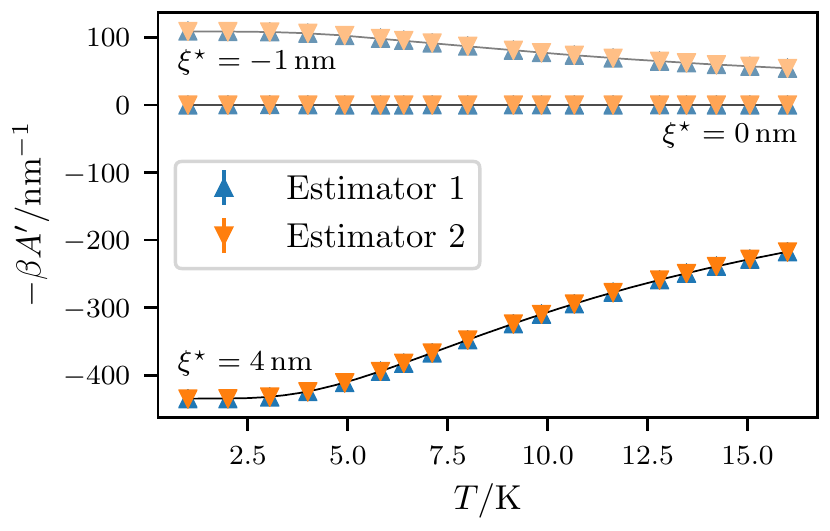}
	\caption{
		Comparison of the estimators $\estim_1$ and $\estim_2$ in Eqs.~\eqref{eq:estim1} and \eqref{eq:estim2} for the computation of the PMF derivative of a harmonic oscillator at $\xi\st = \SI{-1}{\nano\meter}$ (top curve, least saturated), \SI{0}{\nano\meter} (middle curve), and \SI{4}{\nano\meter} (bottom curve, most saturated).
		Error bars are not visible, because they are smaller than the symbols.
		The solid curves show the exact result from Eq.~\eqref{eq:exact-ho1d}.
	}
	\label{fig:results-deriv-ho-1d}
\end{figure}

The necessary quantities for the PIMC estimators are
\begin{subequations}
\begin{align}
	\dlogJ{q}
	&= 0, \\
	\pdv{q}{\xi}
	&= 1, \\
\intertext{and}
	F(q)
	&= -m \omega^2 q.
\end{align}
\end{subequations}
For this example, we have arbitrarily chosen $m = \SI{1.5}{\gram\per\mole}$ and $\omega = \SI{2.3}{\per\pico\second}$.
The derivative of the PMF as computed using the Monte Carlo estimators agrees very well with the exact result over a range of temperatures and constraint positions, as shown in Fig.~\ref{fig:results-deriv-ho-1d}.

\subsection{Lennard-Jones dimer}

To demonstrate that these estimators are applicable to a curvilinear reaction coordinate, we study a diatomic molecule with reduced mass $\mu$ and Lennard-Jones interactions.
Without the term for translation of the center of mass, its Hamiltonian is
\begin{align}
	\label{eq:lj-H}
	\hat{H}
	&= \frac{\hat{p}_{\vec{q}}^2}{2 \mu} + V_\mathrm{LJ}(\hat{\xi}),
\end{align}
where $\vec{q}$ is the radial separation vector between the atoms, whose magnitude $\xi = \abs{\vec{q}}$ we use as the reaction coordinate, and
\begin{align}
	\label{eq:lj-pot}
	V_\mathrm{LJ}(\xi)
	&= 4 \epsilon \left[ \left( \frac{\sigma}{\xi} \right)^{12} - \left( \frac{\sigma}{\xi} \right)^6 \right]
\end{align}
is the Lennard-Jones potential.
Unlike the harmonic oscillator example, this system has a potential that vanishes at large separation, allowing the dimer to dissociate.

Expressing $\vec{q}$ in spherical coordinates $(\xi, \cos{\theta}, \phi)$, we have that the magnitude of the Jacobian determinant is
\begin{align}
	\abs{J(\vec{X}, \xi)}
	&= \xi^2.
\end{align}
In order to evaluate the PIMC estimators, we therefore require the following quantities:
\begin{subequations}
\begin{align}
	\dlogJ{\vec{q}}
	&= \frac{2}{\xi}, \\
	\pdv{\vec{q}}{\xi}
	&= \frac{\vec{q}}{\xi}, \\
\intertext{and}
	\vec{F}(\vec{q})
	&= 24 \epsilon \frac{\vec{q}}{\xi^2} \left[ 2 \left( \frac{\sigma}{\xi} \right)^{12} - \left( \frac{\sigma}{\xi} \right)^6 \right].
\end{align}
\end{subequations}
Note that we retain the geometric term during the simulation and explicitly remove it in the subsequent numerical integration.
To perform a reference calculation, we use numerical matrix multiplication (NMM), as described in Appendix~\ref{sec:nmm-radial}.

\begin{figure}
	\includegraphics{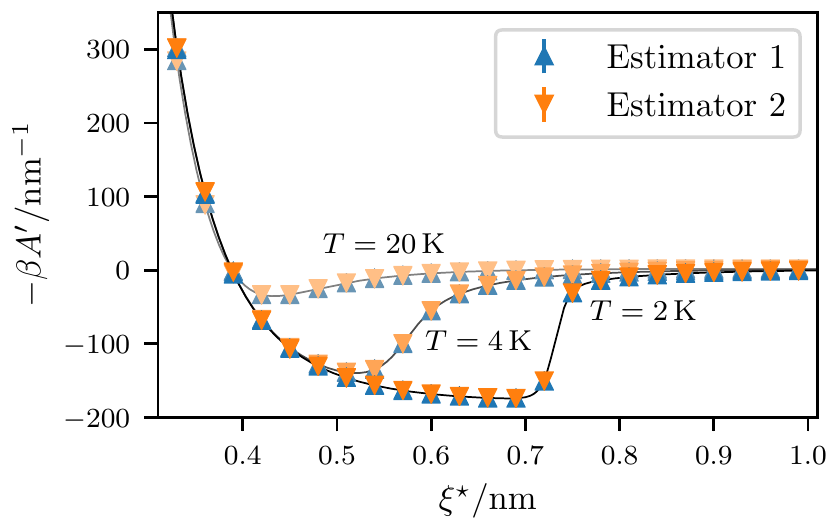}
	\caption{
		Comparison of the estimators $\estim_1$ and $\estim_2$ in Eqs.~\eqref{eq:estim1} and \eqref{eq:estim2} for the computation of the PMF derivative of a Lennard-Jones dimer at $T = \SI{20}{\kelvin}$ (top curve, least saturated), \SI{4}{\kelvin} (middle curve), and \SI{2}{\kelvin} (bottom curve, most saturated).
		Error bars are not visible, because they are smaller than the symbols.
		The solid curves show the NMM results.
	}
	\label{fig:results-deriv-lj3d}
\end{figure}

For the Lennard-Jones parameters provided in Ref.~\onlinecite{thirumalai1983iterative} for \ce{Ar2} ($\epsilon = \SI{119.8}{\kelvin}$ and $\sigma = \SI{3.405}{\angstrom}$), the results in Fig.~\ref{fig:results-deriv-lj3d} confirm that the estimators $\estim_1$ and $\estim_2$ function correctly with radial distance as a reaction coordinate.
In particular, the rapid change in the slope of the PMF is captured at the lower temperatures.

\begin{figure}
	\includegraphics{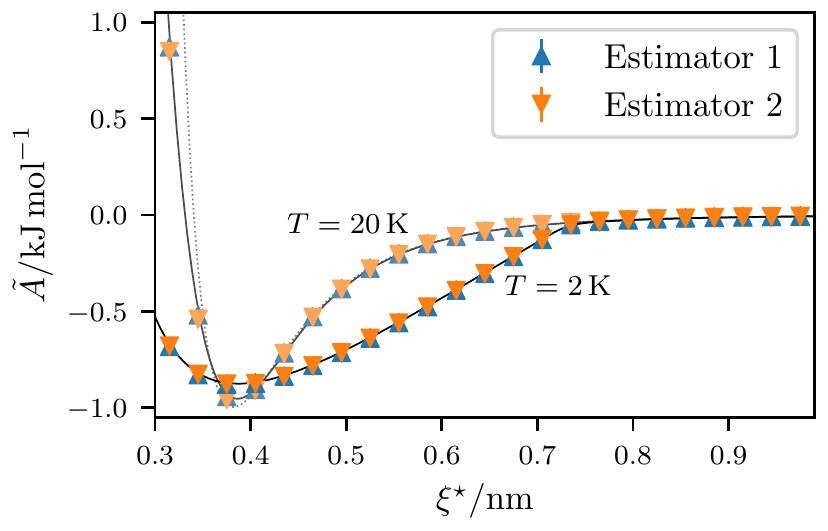}
	\caption{
		Comparison of the estimators $\estim_1$ and $\estim_2$ in Eqs.~\eqref{eq:estim1} and \eqref{eq:estim2} for the computation of the PMF of a Lennard-Jones dimer at $T = \SI{20}{\kelvin}$ (narrow curve, least saturated) and \SI{2}{\kelvin} (wide curve, most saturated).
		Error bars are not visible, because they are smaller than the symbols.
		Additional points extending to $\xi\st_0 = \SI{1.5}{\nano\meter}$ are not displayed.
		The solid curves show the NMM results, while the dotted curve is the Lennard-Jones potential in Eq.~\eqref{eq:lj-pot}.
	}
	\label{fig:results-profile-lj3d}
\end{figure}

It is possible to numerically integrate the derivative $A'(\xi\st)$ to recover the PMF $A(\xi\st)$.
We do so using the midpoint rule on a grid of points $\xi\st_i$ with spacing $\Delta \xi\st$ (as in Fig.~\ref{fig:results-deriv-lj3d}), and with $\xi\st_1$ placed at the largest value of $\xi\st$.
We also define the virtual point $\xi\st_0 = \xi\st_1 + \Delta \xi\st$ and a shifted grid of points
\begin{align}
	\bar{\xi}\st_i
	&= \xi\st_i - \frac{\Delta \xi\st}{2},
\end{align}
with $\bar{\xi}\st_0$ acting as a ``point at infinity'' (the dimer is considered to have dissociated when the atoms are at least $\bar{\xi}\st_0$ apart).
Correspondingly, we set $\tilde{A}(\bar{\xi}\st_0) = 0 = \tilde{A}'(\xi\st_0)$, using the normalization in Eq.~\eqref{eq:spherical-normalization}.

In Fig.~\ref{fig:results-profile-lj3d}, we show
\begin{align}
	\tilde{A}(\bar{\xi}\st_j)
	&= \frac{\Delta \xi\st}{\beta} \sum_{i=1}^j \left[ -\beta A'(\xi\st_i) - \frac{2}{\xi\st_i} \right],
\end{align}
which is the renormalized PMF with the desired energy offset.
The matching NMM curves are calculated from $\rho(\xi\st)$ as
\begin{align}
	\tilde{A}(\xi\st)
	&= -\frac{1}{\beta} \log{\frac{\rho(\xi\st) (\bar{\xi}\st_0)^2}{\rho(\bar{\xi}\st_0) (\xi\st)^2}}
\end{align}
to ensure a compatible energy offset.
Even though the integration grid is rather sparse, especially where the slope of the PMF changes suddenly for $T = \SI{2}{\kelvin}$, the obtained PMFs are consistent with the reference results.

\section{Conclusions}
\label{sec:conclusions}

We have obtained a quantum mechanical expression for the PMF.
This expression is based on the logarithmic derivative of a reduced density operator with respect to a reaction coordinate.
We have provided a path integral representation, and described two PIMC estimators for the calculation of the derivative of the quantum PMF.
Notably, the curves obtained from these estimators are in terms of the true quantum reaction coordinate observable, unlike other methods that utilize the path centroid.

The first estimator, Eq.~\eqref{eq:estim1}, was obtained by initially differentiating the exact path integral and then discretizing the resulting path integral into imaginary time steps.
Alternatively, the second estimator, Eq.~\eqref{eq:estim2}, was obtained by discretizing the exact path integral first and then performing the differentiation after.
In principle, these should be equivalent operations in the $P \to \infty$ limit, and we have demonstrated that both estimators reproduce the correct derivative of the PMF for the one-dimensional harmonic oscillator and Lennard-Jones dimer.
In contrast to existing histogram-based methods for the evaluation of free energies, these novel estimators can be used to ascertain information about the free energy profile at just a single point along the reaction coordinate.

Furthermore, it is possible to numerically integrate the computed derivatives evaluated from these estimators to obtain the PMF itself.
As shown in the argon dimer example, even when the integration grid is not very dense, this method successfully reproduces the known PMF obtained from numerical matrix multiplication.

In Paper II of this series, we show how these estimators may be used with path integral molecular dynamics.
This is achieved by applying techniques from constrained Langevin dynamics to the PILE integrator in order to constrain one of the beads.
The extension of these estimators to path integral molecular dynamics simulations will allow for their application to more general systems and potentials, such as small water clusters.

\begin{acknowledgments}
We thank Raymond Kapral for providing the initial direction for the derivation of the discretized constrained path integral.
This research was supported by the Natural Sciences and Engineering Research Council of Canada (NSERC) (RGPIN-2016-04403), the Ontario Ministry of Research and Innovation (MRI), the Canada Research Chair program (950-231024), and the Canada Foundation for Innovation (CFI) (project No. 35232).
\end{acknowledgments}

\appendix

\section{Kets in curvilinear coordinates}
\label{sec:curvilinear}

A wavefunction $\psi(\vec{q})$ may be thought of as the concrete manifestation of an abstract ket $\ket{\psi}$ in a continuous representation:
\begin{align}
	\psi(\vec{q})
	&= \ip{\vec{q}}{\psi}.
\end{align}
Although the object $\ket{\vec{q}}$ (which represents a state with definite Cartesian position $\vec{q}$) is not an element of Hilbert space, it is common to formally treat it as if it were.
Given a change of variables from $\vec{q}$ to $\vec{X}$, $\xi$ with Jacobian determinant $J(\vec{q}) = J(\vec{X}, \xi)$, it is useful to define $\ket{\vec{X} \, \xi}$ in a way that fulfills
\begin{align}
	\int\! \dd{\vec{q}} \, \abs{\ip{\vec{q}}{\psi}}^2
	&= \int\! \dd{\vec{X}} \int\! \dd{\xi} \, \abs{\ip{\vec{X} \, \xi}{\psi}}^2,
\end{align}
which is analogous to the statement that the resolution of the identity
\begin{align}
	\hat{\mathds{1}}
	&= \int\! \dd{\vec{X}} \int\! \dd{\xi} \, \dyad{\vec{X} \, \xi}
\end{align}
should have the usual form, even in curvilinear coordinates.

Since
\begin{align}
	\int\! \dd{\vec{q}} \, \abs{\ip{\vec{q}}{\psi}}^2
	&= \int\! \dd{\vec{X}} \int\! \dd{\xi} \, \abs{J(\vec{X}, \xi)} \, \abs{\ip{\vec{q}(\vec{X}, \xi)}{\psi}}^2,
\end{align}
it follows that the definition
\begin{align}
	\ket{\vec{X} \, \xi}
	&= \sqrt{\abs{J(\vec{X}, \xi)}} \ket{\vec{q}(\vec{X}, \xi)}
	= \sqrt{\abs{J(\vec{q})}} \ket{\vec{q}}
\end{align}
is sufficient.
This is the approach described in Ref.~\onlinecite{leaf1980curvilinear}, and the one we use in the present work.
Using this definition, we see that the diagonal matrix elements of the partial trace of an operator $\hat{O}$ with respect to $\vec{X}$ may be expressed as
\begin{subequations}
\begin{align}
	\mel{\xi\st}{\Tr_{\vec{X}} \hat{O}}{\xi\st}
	&= \int\! \dd{\vec{X}} \mel{\vec{X} \, \xi\st}{\hat{O}}{\vec{X} \, \xi\st} \\
	&= \int\! \dd{\vec{X}} \int\! \dd{\xi} \, \ddf{\xi - \xi\st} \mel{\vec{X} \, \xi}{\hat{O}}{\vec{X} \, \xi} \\
	&= \int\! \dd{\vec{q}} \, \ddf{\xi(\vec{q}) - \xi\st} \frac{\mel{\vec{X} \, \xi}{\hat{O}}{\vec{X} \, \xi}}{\abs{J(\vec{q})}} \\
	&= \int\! \dd{\vec{q}} \, \ddf{\xi(\vec{q}) - \xi\st} \mel{\vec{q}}{\hat{O}}{\vec{q}}
\end{align}
\end{subequations}
in Cartesian coordinates.

\section{Derivative of a Dirac delta function integral}
\label{sec:delta-derivative}

We wish to take the derivative
\begin{align}
	D(\xi\st)
	&= \dv{\xi\st} \int\! \dd{\vec{q}} \, \ddf{\xi(\vec{q}) - \xi\st} f(\vec{q}).
\end{align}
We first obtain the one-dimensional result
\begin{align}
	\dv{\xi\st} \int\! \dd{\xi} \, \ddf{\xi - \xi\st} f(\xi)
	&= \int\! \dd{\xi} \, \ddf{\xi - \xi\st} \dv{\xi} f(\xi)
\end{align}
by noting that
\begin{align}
	\dv{\xi\st} f(\xi\st)
	&= \int\! \dd{\xi} \, \ddf{\xi - \xi\st} \dv{\xi} f(\xi).
\end{align}
For the general case, we change coordinates to those in which $\xi$ appears explicitly:
\begin{subequations}
\begin{align}
	D(\xi\st)
	&= \int\! \dd{\vec{X}} \dv{\xi\st} \int\! \dd{\xi} \, \ddf{\xi - \xi\st} \abs{J(\vec{X}, \xi)} f(\vec{X}, \xi) \\
	&= \int\! \dd{\vec{X}} \int\! \dd{\xi} \, \ddf{\xi - \xi\st} \pdv{\xi} \abs{J(\vec{X}, \xi)} f(\vec{X}, \xi) \\
	&= \int\! \dd{\vec{X}} \int\! \dd{\xi} \, \ddf{\xi - \xi\st} \left[ \pdv{\xi} \abs{J(\vec{X}, \xi)} \right] f(\vec{X}, \xi)
	\notag \\ &\qquad
		+ \int\! \dd{\vec{X}} \int\! \dd{\xi} \, \ddf{\xi - \xi\st} \abs{J(\vec{X}, \xi)} \pdv{\xi} f(\vec{X}, \xi) \\
	&= \int\! \dd{\vec{q}} \, \ddf{\xi(\vec{q}) - \xi\st} \left[ J_\xi(\vec{q}) + \pdv{\xi} \right] f(\vec{q}),
\end{align}
\end{subequations}
where
\begin{align}
	J_\xi(\vec{q})
	&= \dlogJ{\vec{q}}
	= \frac{
			\pdv{\xi} \abs{J(\vec{X}, \xi)}
		}{
			\abs{J(\vec{X}, \xi)}
		},
\end{align}
and we formally apply the logarithmic derivative notation even when the function is not dimensionless.

\section{Derivative--commutator identity for diagonal matrix elements}
\label{sec:derivative-commutator}

It is well-known that momentum operators lead to differentiation in the position representation.
For example,
\begin{align}
	\mel{\vec{q}}{\hat{p}_i \hat{A}}{\vec{q}'}
	&= -i \hbar \pdv{q_i} \mel{\vec{q}}{\hat{A}}{\vec{q}'}
\end{align}
for an arbitrary operator $\hat{A}$, where $\hat{p}_i$ is the momentum operator conjugate to $\hat{q}_i$.
However, this relationship does not generally hold when $\vec{q}$ and $\vec{q}'$ are the same variable:
\begin{align}
	\mel{\vec{q}}{\hat{p}_i \hat{A}}{\vec{q}}
	&\ne -i \hbar \pdv{q_i} \mel{\vec{q}}{\hat{A}}{\vec{q}}.
\end{align}

Instead, for a Hermitian operator $\hat{A}$ with the eigenvalue equation $\hat{A} \ket{a} = a \ket{a}$, we have that
\begin{subequations}
\begin{align}
	\mel{\vec{q}}{\hat{p}_i \hat{A}}{\vec{q}}
	&= \sum_a \mel{\vec{q}}{\hat{p}_i}{a} \mel{a}{\hat{A}}{\vec{q}} \\
	&= -i \hbar \sum_a a \left[ \pdv{q_i} \ip{\vec{q}}{a} \right] \ip{a}{\vec{q}}
\end{align}
\end{subequations}
and
\begin{align}
	\mel{\vec{q}}{\hat{A} \hat{p}_i}{\vec{q}}
	&= i \hbar \sum_a a \ip{\vec{q}}{a} \left[ \pdv{q_i} \ip{a}{\vec{q}} \right].
\end{align}
Thus, we conclude that
\begin{subequations}
\begin{align}
	\pdv{q_i} \mel{\vec{q}}{\hat{A}}{\vec{q}}
	&= \pdv{q_i} \sum_a \mel{\vec{q}}{\hat{A}}{a} \ip{a}{\vec{q}} \\
	&= \sum_a a \ip{\vec{q}}{a} \left[ \pdv{q_i} \ip{a}{\vec{q}} \right]
	\notag \\ &\qquad
		+ \sum_a a \left[ \pdv{q_i} \ip{\vec{q}}{a} \right] \ip{a}{\vec{q}} \\
	&= \frac{1}{i \hbar} \mel{\vec{q}}{\hat{A} \hat{p}_i}{\vec{q}}
		- \frac{1}{i \hbar} \mel{\vec{q}}{\hat{p}_i \hat{A}}{\vec{q}} \\
	&= \frac{1}{i \hbar} \mel{\vec{q}}{\comm{\hat{A}}{\hat{p}_i}}{\vec{q}}.
\end{align}
\end{subequations}

\section{Numerical matrix multiplication for a radial coordinate}
\label{sec:nmm-radial}

In Ref.~\onlinecite{thirumalai1983iterative}, expressions for numerical matrix multiplication of the path integral of a system described by a three-dimensional relative coordinate are given, but not derived.
In this section, we briefly explain why the radial propagator has such a curious form.

The operator in the kinetic energy of Eq.~\eqref{eq:lj-H} may be expressed as
\begin{align}
	\hat{p}_{\vec{q}}^2
	&= \hat{p}_\xi^2 + \frac{\hat{\ell}^2}{\hat{\xi}^2},
\end{align}
where $\hat{p}_\xi$ is the radial momentum operator, and $\hat{\ell}^2$ is the squared angular momentum operator, whose eigenstates are the spherical harmonics $\ket{\ell \, m}$ with eigenvalues $\hbar^2 \ell (\ell+1)$.
The radial momentum operator is not self-adjoint and does not have a spectrum of eigenstates,\cite{liboff1973radial,paz2002non} so the spectral theorem does not apply to it and the appropriate resolution of identity is not given by
\begin{align}
	\int\! \dd{p_\xi} \dyad{p_\xi},
\end{align}
despite the wavefunctions
\begin{align}
	\ip{\xi}{p_\xi}
	&= \frac{e^{\frac{i \xi p_\xi}{\hbar}}}{\sqrt{2\pi \hbar}}
\end{align}
satisfying $\hat{p}_\xi \ket{p_\xi} = p_\xi \ket{p_\xi}$.
Thus, we must be careful when rederiving Eq.~(16c) of Ref.~\onlinecite{thirumalai1983iterative}.

We turn to the operator $\hat{p}_\xi^2$, which is well-behaved and has the eigenstates
\begin{align}
	\ip*{\xi}{p_\xi^{(2)}}
	&= \frac{e^{\frac{i \xi p_\xi}{\hbar}} - e^{\frac{-i \xi p_\xi}{\hbar}}}{2 i \sqrt{\pi \hbar}}
	= \frac{1}{\sqrt{\pi \hbar}} \sin{\frac{\xi p_\xi}{\hbar}}
\end{align}
with eigenvalues $p_\xi^2$.
We may use these states to construct the correct resolution of the identity,
\begin{align}
	\hat{\mathds{1}}
	&= \int\! \dd{p_\xi} \dyad*{p_\xi^{(2)}}
	= \int\! \dd{p_\xi} \Big( \dyad{p_\xi} - \dyad{p_\xi}{-p_\xi} \Big),
\end{align}
which, as expected, results in
\begin{widetext}
\begin{subequations}
\begin{align}
	\mel{\xi'}{e^{-\frac{\tau \hat{p}^2_\xi}{2 \mu}}}{\xi}
	&= \int\! \dd{p_\xi} \mel*{\xi'}{e^{-\frac{\tau p_\xi^2}{2 \mu}}}{p_\xi^{(2)}} \ip*{p_\xi^{(2)}}{\xi} \\
	&= \frac{1}{4\pi \hbar} \int\! \dd{p_\xi} e^{-\frac{\tau p_\xi^2}{2 \mu} + \frac{i p_\xi}{\hbar} (\xi' - \xi)}
		+ \frac{1}{4\pi \hbar} \int\! \dd{p_\xi} e^{-\frac{\tau p_\xi^2}{2 \mu} - \frac{i p_\xi}{\hbar} (\xi' - \xi)}
	\notag \\ &\qquad
		- \frac{1}{4\pi \hbar} \int\! \dd{p_\xi} e^{-\frac{\tau p_\xi^2}{2 \mu} + \frac{i p_\xi}{\hbar} (\xi' + \xi)}
		- \frac{1}{4\pi \hbar} \int\! \dd{p_\xi} e^{-\frac{\tau p_\xi^2}{2 \mu} - \frac{i p_\xi}{\hbar} (\xi' + \xi)} \\
	&= \sqrt{\frac{\mu}{2\pi \hbar^2 \tau}} \left[
				e^{-\frac{\mu}{2 \hbar^2 \tau} (\xi' - \xi)^2}
				- e^{-\frac{\mu}{2 \hbar^2 \tau} (\xi' + \xi)^2}
			\right].
\end{align}
\end{subequations}
\end{widetext}

\end{document}